\documentclass[12pt]{article}%
\usepackage{amsmath,latexsym}
\usepackage{graphicx}
\usepackage{amsmath}
\usepackage{amsfonts}
\usepackage{amssymb}%
\begin{document}

\title{{On the status of wormholes in Einstein's
   theory: an overview}}
   \author{
Peter K. F. Kuhfittig*\\  \footnote{kuhfitti@msoe.edu}
 \small Department of Mathematics, Milwaukee School of
Engineering,\\
\small Milwaukee, Wisconsin 53202-3109, USA}

\date{}
 \maketitle

\begin{abstract}\noindent
It has been claimed that wormholes are just
as good a prediction of Einstein's theory as
black holes, but they are subject to severe
restrictions from quantum field theory.  The
purpose of this paper is to show that the claim
can be substantiated in spite of these
restrictions.   
\\

\end{abstract}

\section{Introduction}\label{E:introduction}

The laws of physics are often used to make
solid inferences.  For example, Newton's
laws allow the determination of the motion
of a weight hanging on a spring. In other
situations, the laws may simply allow something
to happen.  For example, Einstein's theory
allows backward time travel but does not imply
that backward time travel can actually be
achieved.  Similar comments can be made about
macroscopic traversable wormholes: while
wormholes are just as good a prediction of
Einstein's theory as black holes, they are
subject to severe restrictions from quantum
field theory.  An example is the need to
violate the null energy condition, calling
for the existence of ``exotic matter" (defined
below) to hold a wormhole open.  Its
problematical nature has caused many researchers
to consider such wormholes solutions to be
completely unphysical.

The continuing interest in wormholes is based
on the observation that the Schwarzschild
solution and therefore black holes describe a
(nontraversable) wormhole.  More recent
developments involving entanglement have
suggested that a special type of wormhole,
called an Einstein-Rosen bridge, may be the
best explanation for entanglement \cite{MS13}.
We will therefore assume that a basic
wormhole structure can be hypothesized.

While there had been some forerunners,
macroscopic traversable wormholes were first
studied by Morris and Thorne \cite{MT88}, who
proposed the following static and spherically
symmetric line element for a wormhole spacetime:
\begin{equation}\label{E:line1}
ds^{2}=-e^{2\Phi(r)}dt^{2}+e^{2\alpha(r)}dr^2
+r^{2}(d\theta^{2}+\text{sin}^{2}\theta\,
d\phi^{2}),
\end{equation}
where
\begin{equation}
   e^{2\alpha(r)}=\frac{1}{1-\frac{b(r)}{r}}.
\end{equation}
(We are using units in which $c=G=1$.)  In the now
customary terminology, $\Phi=\Phi(r)$ is called
the \emph{redshift function}, which must be finite
everywhere to prevent the occurrence of an event
horizon. The function $b=b(r)$ is called the
\emph{shape function} since it determines the
spatial shape of the wormhole when viewed, for
example, in an embedding diagram \cite{MT88}.
The spherical surface $r=r_0$ is the \emph{throat}
of the wormhole.  In a Morris-Thorne wormhole, the
shape function must satisfy the following
conditions: $b(r_0)=r_0$, $b(r)<r$ for $r>r_0$,
and $b'(r_0)<1$, called the \emph{flare-out
condition} in Ref. \cite{MT88}.  In classical
general relativity, the flare-out condition can
only be met by violating the null energy condition
(NEC), which states that for the energy-momentum
tensor $T_{\alpha\beta}$
\begin{equation}
   T_{\alpha\beta}k^{\alpha}k^{\beta}\ge 0\,\,
   \text{for all null vectors}\,\, k^{\alpha}.
\end{equation}
Matter that violates the NEC is called ``exotic"
in Ref. \cite{MT88}.  Applied to a wormhole
setting, observe that for the radial outgoing
null vector $(1,1,0,0)$, the violation reads
\begin{equation}
   T_{\alpha\beta}k^{\alpha}k^{\beta}=\rho+
      p_r<0.
\end{equation}
  Here $T^t_{\phantom{tt}t}=-\rho(r)$
is the energy density, $T^r_{\phantom{rr}r}=
p_r(r)$ is the radial pressure, and
$T^\theta_{\phantom{\theta\theta}\theta}=
T^\phi_{\phantom{\phi\phi}\phi}=p_t(r)$
is the lateral (transverse) pressure.  Our
final requirement is \emph{asymptotic
flatness:}
\begin{equation}
   \text{lim}_{r\rightarrow\infty}\Phi(r)=0
   \quad \text{and} \quad
   \text{lim}_{r\rightarrow\infty}
   \frac{b(r)}{r}=0.
\end{equation}

The problematical nature of exotic matter
in conjunction with the need to violate
the NEC has suggested solutions beyond the
classical theory.  For example, it was
proposed by Lobo and Oliveira \cite{LO09}
that in $f(R)$ modified gravity, the
wormhole throat could be lined with
ordinary matter, while the violation of
the NEC can be attributed to the higher-order
curvature terms.  There exist a number of
other modified theories of gravity that
could be called upon to address these issues.

The primary goal of this paper is to
accommodate the energy violation without
modifying Einstein's theory.

\section{Meeting the goal: preliminaries}
   \label{S:goal}
When dealing with a complex theory such as
general relativity, certain aspects can
be viewed from a broader perspective that
stops short of a modification.  For
example, the extension of Einstein's
theory to higher dimensions has a  long
history, eventually leading to the
realization that Einstein's theory
is the low-energy limit of string
theory (with its extra dimensions), 
just as Newton's theory is the
weak-gravity and low-velocity
limit of Einstein's theory.  To be 
consistent with our goal, we will 
consider an extra dimension to be 
a natural extension of Einstein's 
theory, rather than a modification.  
So much of our interest is going
to be centered on Refs. \cite{pK18}
and \cite{pK19}, which hypothesize
an extra static and time-dependent
spatial dimension, respectively.
These topics are covered in Sections
\ref{S:static}-\ref{S:time}.

Another striking development is
noncommutative geometry, which may
be viewed as another offshoot of
string theory.  As described in Sec.
\ref{S:emergent}, point-like particles
are replaced by smeared objects, which
is consistent with Heisenberg's
uncertainty principle.  What is
critically important for our purposes
is that the noncommutative effects
can be implemented in the Einstein
field equations by modifying only
the energy momentum tensor while
leaving the Einstein tensor intact,
once again avoiding a modification
of Einstein's theory.  These topics
are discussed in Sections
\ref{S:emergent} and \ref{S:Casimir}.

The realization that moderately-sized
wormholes are subject to an enormous
radial tension suggests that wormholes
are actually compact stellar objects.
A possible explanation is sought in
Sec. \ref{S:neutron} by starting with
a two-fluid model that was previously
proposed in Ref. \cite{fR12}.  Additional
assumptions are unavoidable, however, as
we will see in Sec. \ref{S:neutron}.

Exotic matter makes a brief comeback in Sec.
\ref{S:revisit}.  Small amounts that may
arise from, for example, the Casimir effect,
call for striking a delicate balance between
reducing the amount of exotic matter and
fine-tuning the metric coefficients.

Finally, it is shown in Sec. \ref{S:stability}
that a noncommutative-geometry wormhole in a
static and spherically symmetric spacetime
admitting conformal motion is stable to
linearized radial perturbations.  Furthermore,
both the redshift and shape functions are
completely determined from the given
conditions.

In Sec. \ref{S:other} we take another, more 
general, look at the low energy density in a
noncommutative-geometry setting by comparing the
outcome to other low-density models, including
the case $\rho(r)\equiv 0$; none of these have
the special characteristics of the former.
Without these special features, the need for
exotic matter cannot be avoided, indicating
that neither dark matter nor dark energy can
support traversable wormholes, at least not
as long as the latter does not cross the
phantom divide.

Sec. \ref{S:lensing} discusses the possible
detection of wormholes by means of gravitational
lensing.  This tool calls for additional physical
requirements beyond the existence of dark matter,
thereby confirming the above assertion.

\section{An extra spatial dimension (static case)}
   \label{S:static}
In this section, our main interest will be centered
on Ref. \cite{pK18}, which involves an extra spatial
dimension.  The extended line element is
\begin{equation}\label{E:line2}
ds^{2}=-e^{2\Phi(r)}dt^{2}+\frac{dr^2}{1-\frac{b(r)}{r}}
+r^{2}(d\theta^{2}+\text{sin}^{2}\theta\,
d\phi^{2})+e^{2\mu(r,l)}dl^2,
\end{equation}
where $l$ is the extra coordinate.

It is interesting to note that in the
components of the Riemann curvature tensor,
$\mu(r,l)$ never occurs as a factor.  Instead,
the factors are $\partial\mu(r,l)/\partial r$
and $\partial^2\mu(r,l)/\partial r^2$
\cite{pK18}.  So $\mu(r,l)$ could have any
magnitude, to be discussed further below.

According to Ref. \cite{pK18},
\begin{equation}
  (\rho +p_r)|_{r=r_0}=\frac{1}{8\pi}\frac{b'(r_0)-1}{r_0^2}
  \left[1+\frac{r_0}{2}\frac{\partial\mu(r_0,l)}{\partial r}
  \right].
\end{equation}
To satisfy the condition $\rho +p_r>0$ at the throat,
we must have
\begin{equation}\label{E:condition1}
   \frac{\partial\mu(r_0,l)}{\partial r}<-\frac{2}{r_0},
\end{equation}
corresponding to the null vector $(1,1,0,0)$.  Moving
to the fifth dimension, the null vector $(1,0,0,0,1)$
yields
 \begin{equation}\label{E:condition2}
   (\rho +p_r)|_{r=r_0}\left.=\frac{1}{8\pi}
    \frac{1}{2}\frac{rb'-b}{r^2}\left[
    -\frac{d\Phi(r)}{dr}
    +\frac{\partial\mu(r,l)}{\partial r}
    \right]\right|_{r=r_0}<0,
\end{equation}
provided that the redshift function satisfies a
similar condition:
\begin{equation}\label{E:condition3}
   \frac{d\Phi(r_0)}{dr}=-A<
   \frac{\partial\mu(r_0,l)}{\partial r}
   <-\frac{2}{r_0}.
\end{equation}
We conclude that the NEC is satisfied at
the throat in the four-dimensional spacetime
but violated in the five-dimensional spacetime.

\emph{Remark 1:} For the condition
$\rho(r_0)+p_r(r_0)>0$ to hold for all null
vectors, we must also have $b'(r_0)>1/3$
\cite{pK18}.

\emph{Remark 2:} Condition (\ref{E:condition3})
can be readily satisfied if $\Phi=\Phi(r)$ is
a positive differentiable decreasing function
of $r$ for all $r$.  The reason is that since
$\Phi'(r)<0$, the assumption of asymptotic
flatness implies that
$\text{lim}_{r\rightarrow\infty}\Phi'(r)=0$.

\section{Additional considerations}

\subsection{The function $\mu=\mu(r,l)$}
   \label{S:extra}
We have already seen that Inequality
(\ref{E:condition1}) is a sufficient
condition for ensuring that the throat of a
wormhole can be threaded with ordinary matter,
while the unavoidable violation of the NEC
can be attributed to the higher spatial
dimension.

Here we need to emphasize another aspect of
$\mu(r,l)$: as noted in the previous section,
$|\mu(r,l)|$ can be large or small.  So in
our model, it is entirely possible that
$\mu$ be negative with a large absolute
value, resulting in a small value for
$e^{\mu(r,l)}$.  In other words, the extra
dimension could be compactified without
sacrificing Inequality (\ref{E:condition1}).
The very existence of a compactified extra
dimension is consistent with string theory.
So the assumptions regarding $\mu(r,l)$ are
physically reasonable.

Summarizing the static case, $\Phi=\Phi(r)$
is positive and decreasing, while $b'(r_0)
>1/3$.  Conditions (\ref{E:condition1}) and
(\ref{E:condition3}) are physically
reasonable and consistent with string
theory.

\subsection{The radial tension at the throat}
   \label{S:tension}
At this point, we need to return to Ref.
\cite{MT88} to discuss the radial tension
at the throat.  To that end, we need to
recall that the radial tension $\tau(r)$
is the negative of the radial pressure
$p_r(r)$.  According to Ref. \cite{MT88},
the Einstein field equations can be
rearranged to yield $\tau(r)$.  Here we
need to reintroduce $c$ and $G$ temporarily
to get
\begin{equation}
   \tau(r)=\frac{b(r)/r-2[r-b(r)]\Phi'(r)}
   {8\pi Gc^{-4}r^2}.
\end{equation}
So the radial tension at the throat becomes
\begin{equation}\label{E:tau}
  \tau(r_0)=\frac{1}{8\pi Gc^{-4}r_0^2}\approx
   5\times 10^{41}\frac{\text{dyn}}{\text{cm}^2}
   \left(\frac{10\,\text{m}}{r_0}\right)^2.
\end{equation}
As pointed out in Ref. \cite{MT88}, for
$r_0=3$ km, $\tau(r)$ has the same magnitude
as the pressure at the center of a massive
neutron star.  (For further discussion of
this problem, see Ref. \cite{pK22}.)  So
it follows from Eq. (\ref{E:tau}) that
wormholes with a low radial tension could
only exist on very large scales.

\section{An extra spatial dimension
 (time-dependent case)}\label{S:time}

So far the wormhole geometry has been strictly
static.  Ref. \cite{pK19} discusses the case
in which the extra dimension is a function of
time $t$, as well  as $r$ and $l$.  If the
shape and redshift functions remain the same,
then the line element becomes
\begin{equation}\label{E:line3}
ds^{2}=-e^{2\Phi(r)}dt^{2}+\frac{dr^2}{1-\frac{b(r)}{r}}
+r^{2}(d\theta^{2}+\text{sin}^{2}\theta\,
d\phi^{2})+e^{2\mu(r,l,t)}dl^2.
\end{equation}
(Ref. \cite{pK19} also assumes that $b$ and
$\Phi$ are functions of $r$ and $l$.)  The
expression for the null energy condition
is given by
\begin{multline}\label{E:NEC}
   8\pi(\rho +p_r)|_{r=r_0}=\frac{b'(r_0)-1}{r_0^2}\
   \left[1+\frac{r_0}{2}\frac{\partial\mu(r_0,l,t)}
   {\partial r}\right]\\
   -e^{-2\Phi(r_0)}\left[
   \frac{\partial^2\mu(r_0,l,t)}{\partial t^2}
   +\left(\frac{\partial\mu(r_0,l,t)}
   {\partial t}\right)^2\right].
\end{multline}
To put this result to use, we can start with
Ref. \cite{pK23b}, which deals with a wormhole
model due to S.-W. Kim \cite{sK96} in conjunction
with a generalized Kaluza-Klein model:
\begin{equation}\label{E:L4}
  ds^2=-e^{2\Phi(r)}dt^2+[a(t)]^2\left(\frac{dr^2}
  {1-kr^2-\frac{b(r)}{r}}+r^2
  (d\theta^2+\text{sin}^2\theta\,d\phi^2)
  +e^{2\Psi(r)}dq^2\right).
\end{equation}
However, our primary concern is the effect of the
time-dependent extra dimension, rather than the
overall cosmological model.  So we can let
$k=0$ and assume that the line element has the
form
\begin{equation}\label{E:L5}
  ds^2=-e^{2\Phi(r)}dt^2+\frac{dr^2}
  {1-\frac{b(r)}{r}}+r^2
  (d\theta^2+\text{sin}^2\theta\,d\phi^2)
  +[a(t)]^2e^{2\mu(r,l)}dl^2,
\end{equation}
using our earlier notation for the last term.
Since $[a(t)]^2e^{2\mu(r,l)}=
e^{2(\text{ln}\,a(t)+\mu(r,l))}$,
we let
\begin{equation}
   U=\text{ln}\,a(t)+\mu(r,l).
\end{equation}
Then the line element becomes
\begin{equation}\label{E:L6}
  ds^2=-e^{2\Phi(r)}dt^2+\frac{dr^2}
  {1-\frac{b(r)}{r}}+r^2
  (d\theta^2+\text{sin}^2\theta\,d\phi^2)
  +e^{2U(r,l,t)}dl^2.
\end{equation}
So we can now rewrite Eq. (\ref{E:NEC})
in the form
\begin{multline}\label{E:null}
   8\pi(\rho +p_r)|_{r=r_0}=
   \frac{b'(r_0)-1}{r_0^2}\left[1+
   \frac{r_0}{2}\frac{\partial U(r_0,l,1)}
   {\partial r}\right]\\
    -e^{-2\Phi(r_0)}\left[
   \frac{\partial^2U(r_0,l,t)}{\partial t^2}+
   \left(\frac{\partial U(r_0,l,t)}{\partial t}
   \right)^2\right].
\end{multline}
We then observe that
\begin{equation}
   \frac{\partial^2U}{\partial t^2}+
   \left(\frac{\partial U}{\partial t}\right)^2
   =\frac{a''(t)}{a(t)}.
\end{equation}
There are now two possibilities, $a''(t)<0$
and $a''(t)>0$ for the non-static case.  In
the first case, the second term on the
right-hand side of Eq. (\ref{E:null}) is
positive.  It now becomes apparent that
Inequality
(\ref{E:condition1}) can be replaced by
the slightly more general
\begin{equation}
   \frac{\partial\mu(r_0,l,t)}{\partial
   r}\le -\frac{2}{r_0}
\end{equation}
since
\begin{equation*}
  \frac{\partial U}{\partial r}=
  \frac{\partial \mu}{\partial r}.
\end{equation*}
In other words, from Eq. (\ref{E:null}),
we now have
\begin{equation}
   8\pi(\rho +p_r)|_{r=r_0}>0,
\end{equation}
as in the static case.

If $a''(t)>0$, then
\begin{equation}
   8\pi(\rho +p_r)|_{r=r_0}<0,
\end{equation}
and we are back to exotic matter, the case
that was previously dismissed as unphysical.\

\emph{Remark 3:} It is shown in Ref.
\cite{pK19} that the NEC is violated in the
five-dimensionsl spacetime, as in the static
case.

\section{Macroscopic wormholes as emergent phenomena}
   \label{S:emergent}
In this section, our study of wormholes will
move in a different direction by starting
with noncommutative geometry.  Sometimes
viewed as an offshoot of string theory, it
assumes that point-like particles are
replaced by smeared objects, which is
consistent with the Heisenberg uncertainty
principle.  The original idea was to
eliminate the divergences that normally
occur in general relativity \cite{SS03,
NSS06, NS10}.  According to Ref. \cite{NSS06},
this objective can be met by asserting that
spacetime can be encoded in the commutator
$[\textbf{x}^{\mu},\textbf{x}^{\nu}]
=i\theta^{\mu\nu}$, where $\theta^{\mu\nu}$ is
an antisymmetric matrix that determines the
fundamental cell discretization of spacetime
in the same way that Planck's constant $\hbar$
discretizes phase space.  More concretely, the
smearing can be modeled by using a so-called
Lorentzian distribution of minimal length
$\sqrt{\beta}$ instead of the Dirac delta
function \cite{NM08, LL12}.  As a consequence,
the energy density of a static and spherically
symmetric and particle-like gravitational
source is given by
\begin{equation}\label{E:density}
   \rho(r)=\frac{m\sqrt{\beta}}{\pi^2
   (r^2+\beta)^2}.
\end{equation}
The implication is that the gravitational
source causes the mass $m$ to be diffused
throughout the region of linear dimension
$\sqrt{\beta}$ due to the uncertainty.

Returning to line element (\ref{E:line1}),
let us list the Einstein field equations
next:
\begin{equation}\label{E:Ein1}
  \rho(r)=\frac{b'}{8\pi r^2},
\end{equation}
\begin{equation}\label{E:Ein2}
   p_r(r)=\frac{1}{8\pi}\left[-\frac{b}{r^3}+
   2\left(1-\frac{b}{r}\right)\frac{\Phi'}{r}
   \right],
\end{equation}
and
\begin{equation}\label{E:Ein3}
   p_t(r)=\frac{1}{8\pi}\left(1-\frac{b}{r}\right)
   \left[\Phi''-\frac{b'r-b}{2r(r-b)}\Phi'
   +(\Phi')^2+\frac{\Phi'}{r}-
   \frac{b'r-b}{2r^2(r-b)}\right].
\end{equation}
Eq. (\ref{E:density}) now provides a physical
basis for checking the NEC, i.e.,
\begin{multline}\label{E:null2}
     T_{\alpha\beta}k^{\alpha}k^{\beta}=
     \rho(r)+p_r(r)=\frac{m\sqrt{\beta}}
  {\pi^2(r^2+\beta)^2}+\left.\frac{1}{8\pi}
  \left[-\frac{b}{r^3}+2\left(1-\frac{b}{r}
  \right)\frac{\Phi'}{r}\right]\right|
  _{r=r_0}\\
  =\frac{m\sqrt{\beta}}
  {\pi^2(r_0^2+\beta)^2}-\frac{1}{8\pi}
  \frac{b(r_0)}{r_0^3}<0
\end{multline}
since $\sqrt{\beta}\ll m$.  So the violation
of the NEC can be attributed to the
noncommutative-geometry background, rather
than some hypothetical ``exotic matter,"
\emph{at least locally}.  (We will return
to this point at the end of the section.)

For our purposes, it is sufficient to note
that, according to Ref. \cite{pK23a},  the
shape function $B$ is given by
 \begin{multline}
     B\left(\frac{r}{\sqrt{\beta}}\right)=\\
   \frac{4m}{\pi}\frac{1}{r}\left[\frac{r}{\sqrt{\beta}}
   \,\text{tan}^{-1}\frac{r}{\sqrt{\beta}}
   -\frac{\left(\frac{r}{\sqrt{\beta}}\right)^2}
   {\left(\frac{r}{\sqrt{\beta}}\right)^2+1}
  -\frac{r}{\sqrt{\beta}}\,
  \text{tan}^{-1}\frac{r_0}{\sqrt{\beta}}
  +\frac{r}{\sqrt{\beta}}
  \frac{\frac{r_0}{\sqrt{\beta}}}
  {\left(\frac{r_0}{\sqrt{\beta}}\right)^2+1}
  \right]+\frac{r_0}{\sqrt{\beta}}
\end{multline}
and meets all the requirements of a shape
function.  In particular,
\begin{equation}\label{E:B}
   B\left(\frac{r_0}{\sqrt{\beta}}\right)=
   \frac{r_0}{\sqrt{\beta}},
 \end{equation}
which corresponds to $b(r_0)=r_0$.  It follows
that the throat radius is macroscopic.  (See
Ref. \cite{pK23a} for details.)

This outcome naturally raises the question
whether a modification of Einstein's theory
has really been avoided.  It is argued in
Ref. \cite{pK23a} that noncommutative
geometry in the form discussed above is a
\emph{fundamental property} and that the
outcome, a macroscopic throat size, is an
\emph{emergent property}.  By definition,
emergent phenomena are derived from some
fundamental theory, an idea that dates at
least from the time of Aristotle.  For
example, life emerges from totally lifeless
objects, such as atoms and molecules.  This
process is not reversible: living organisms
tell us little about the particles in the
fundamental theory.  Similarly, our emerging
macroscopic scale does not yield the smearing
effect in the fundamental theory.  In the
usual terminology, we have obtained an
\emph{effective model} for a macroscopic
wormhole in the sense that the short-distance
effects have been discarded: these are
meaningful only in the fundamental theory.
The above local violation of the NEC,
$T_{\alpha\beta}k^{\alpha}k^{\beta}<0$,
can therefore be viewed as a fundamental
property.  So the emergent macroscopic
phenomenon in Eq. (\ref{E:B}) avoids a
modification of Einstein's theory.

\section{Neutron stars}\label{S:neutron}
With Sec. \ref{S:tension} in mind, wormholes
should be viewed as compact stellar objects.
The reason is that $\tau(r)$ has the same
magnitude as the pressure at the center of
a massive neutron star.  Moreover, Eq.
(\ref{E:tau}) implies that a wormhole with
a low radial tension could only exist on a
very large scale, i.e., with a sufficiently
large $r=r_0$.  According to Ref.
\cite{pK22c}, for smaller wormholes, even
the boundary condition $b(r_0)=r_0$ only
makes sense if the wormhole is a compact
stellar object.

It is interesting to note that a combined
model consisting of neutron-star matter and
a phantom/ghost scalar field yields a
wormhole solution \cite{vD12}.  Another
example of a two-fluid model can be found
in Ref. \cite{fR12}.  For this approach to
work, we need to follow Ref. \cite{pK13}
which assumes that quark matter exists
at the center of neutron stars.  While 
this may seem like a strong assumption, 
it is by no means 
unreasonable: the extreme conditions could
presumably cause the neutrons to become
deconfined, resulting in quark matter.
Armed with this assumption, the energy
momentum tensor of the two-fluid model is
given by \cite{fR12}
\begin{equation}\label{E:rho}
   T^0_0\equiv\rho_{\text{effective}}
       =\rho+\rho_q,
\end{equation}
\begin{equation}\label{E:p}
   T^1_1=T^2_2\equiv-p_{\text{effective}}
       =-(p+p_q).
\end{equation}
Here $\rho$ and $p$ correspond to the
respective energy density and pressure
of the baryonic matter, while $\rho_q$
and $p_q$ correspond to the respective
energy density and pressure of the quark
matter.  The left-hand sides are the
effective energy density and pressure,
respectively, of the combination.

 The two-fluid model is based on
the MIT bag model \cite{aC74}.  In this
model, the equation of state is given by
\begin{equation}\label{E:bag}
  p_q=\frac{1}{3}(\rho_q-4B),
\end{equation}
where $B$ is the bag constant, which is
given as $145\,\text{MeV}/(\text{fm})^3$
in Ref. \cite{aC74}.  For normal matter,
we can use the rather idealized equation
of state \cite{RKR}
\begin{equation}\label{E:EoS}
   p=m\rho, \quad 0<m<1.
\end{equation}
For our purposes, it is sufficient to note
that Ref. \cite{pK13}  gives the following
solution:
\begin{equation}\label{E:rho1}
  \rho=\rho_0e^{-\Phi(1+m)/2m}
\end{equation}
and
\begin{equation}\label{E:rho2}
    \rho_q=B+\rho_{(q,0)}e^{-2\Phi},
\end{equation}
where $\rho_0$ and $\rho_{(q,0)}$ are
integration constants.  Ref. \cite{pK13}
then goes on to derive the shape function
$b=b(r)$, as well as its derivative
\begin{equation}
   b'(r)=1-e^{-\alpha(r)}+r\left[
   -\frac{d}{dr}e^{-\alpha(r)}
   \right],
\end{equation}
where $e^{\alpha(r)}=1/(1-b(r)/r).$  (See
Ref. \cite{pK13} for details.)  It is
subsequently shown that the flare-out
condition is met, indicating a violation
of the null energy condition, a  necessary
condition for the existence of wormholes.
This violation can be attributed to the
extreme conditions at the center of
neutron stars.

\section{Exotic matter revisited}
     \label{S:revisit}
\subsection{Fine-tuning}\label{S:finetune}

Given our main goal, demonstrating the
possible existence of wormholes without
modifying Einstein's theory, it seems
surprising that an earlier attempt by
the author required no more than
sufficient fine-tuning of the metric
coefficients \cite{pK08}.

First we need to recall that in classical
general relativity, a wormhole can only
be held open by violating the NEC,
calling for the need for exotic matter.
Such matter is confined to a small region
around the throat.  By itself, this is
not a conceptual problem, as showm by the
Casimir effect \cite{pK23c}, to be
discussed further in the next section.
In other words, exotic matter can be
made in the laboratory, but only in small
quantities that may not be sufficient
for keeping a wormhole open.  One of the
goals in Ref. \cite{pK08} is to strike a
balance between two conflicting
requirements, reducing the amount of
exotic matter and fine-tuning the values
of the metric coefficients.

The key to the problem is the discovery
by Ford and Roman that quantum field
theory places severe constraints on the
wormhole geometries \cite{FR95, FR06}.
Of particular interest to us is Eq. (95)
in Ref. \cite{FR06}:
\begin{equation}\label{E:FR}
   \frac{r_m}{r}\le \left(\frac{1}{v^2-b'(r_0)}
   \right)^{1/4}\frac{\sqrt{\delta}}{f}
   \left(\frac{l_p}{r_0}\right)^{1/2},
\end{equation}
where $\delta=1/\sqrt{1-v^2}$, $l_p$ is
the Planck length, $f$ is a small scale
factor, $b'_0=b'(r_0)$, and $r_m$ is the
smallest of several length scales:
\begin{equation}
   r_m\equiv \text{min}\left[b(r),
   \left|\frac{b(r)}{b'(r)}\right|,
   \frac{1}{|\Phi'(r)|}, \left|\frac{\Phi'(r)}
   {\Phi''(r)}\right|
   \right],
\end{equation}
referring once again to line element
(\ref{E:line1}).  Finally, $v$ is the
velocity of a boosted observer relative
to a static frame.  For the right-hand
side of Inequality (\ref{E:FR}) to be
defined and real, we must have $v^2>b'_0$.
So if $b'_0\approx 1$, the inequality is
trivially satisfied, thereby meeting the
Ford-Roman constraints.  However, to
study the region away from the throat,
where $b'(r_0)<1$, Inequality
(\ref{E:FR}) has to be extended, as we
will see shortly.

Before continuing, we need to take a
closer look at the exotic region
\begin{equation}
   l(r)=\int^r_{r_0}e^{\alpha(r')}dr'.
\end{equation}
So $l(r_1)$ is the amount of exotic
matter in the interval $[r_0,r_1]$.
(This is a more precise way of
saying that the exotic matter is
confined to the spherical shell of
inner radius $r=r_0$ and outer
radius $r=r_1$.)

Now consider the extended quatum
inequality from Ref. \cite{pK08}:
\begin{equation}\label{E:pK}
   \frac{r_m}{r}\le \left(\frac{1}
   {v^2\frac{b(r)}{r}-b'(r)-2v^2\Phi'(r)
   \left(1-\frac{b(r)}{r}\right)}\right)^{1/4}
   \frac{\sqrt{\delta}}{f}
   \left(\frac{l_p}{r}\right)^{1/2}.
\end{equation}
At $r=r_0$, Inequality (\ref{E:pK})
reduces to Inequality (\ref{E:FR}).  As a
result,  we are still interested in the
case where $b'(r_0)$ is close to unity
because this leads to our main result:
since $b'(r)<1$, for $r>r_0$, Inequality
(\ref{E:FR}) is not necessarily satisfied,
but in the extended Inequality (\ref{E:pK}),
$\Phi'(r)$ can be fine-tuned so that the
condition is satisfied in the interval
$[r_0,r_1]$, thereby reducing the proper
thickness of the exotic region, perhaps
indefinitely.  (Ref. \cite{pK08} discusses
additional models and gives several
numerical estimates.)

The conclusion is that one must strike a
balance between the thickness $[r_0,r_1]$
of the exotic region and the degree of
fine-tuning required to achieve this
reduction.  It is also shown in Ref.
\cite{pK08} that the degree of
fine-tuning is a generic feature of
a Morris-Thorne wormhole.  This
unexpected finding could be viewed
as an engineering challenge that
some day might even be met.

\subsection{The Casimir effect and
   noncommutative geometry}
        \label{S:Casimir}
The Casimir effect mentioned in Sec.
\ref{S:finetune} has shown that exotic
matter can exist on a small scale.  While
this may not be enough to guarantee that
the Casimir effect can support a
macroscopic wormhole, the fine-tuning
scheme in Sec. \ref{S:finetune} seems
to allow such a possibility.  Another
possibility is discussed in Ref.
\cite{pK23c}: the Casimir effect can
be connected to noncommutative geometry,
which also deals with small-scale effects,
as discussed in Sec. \ref{S:emergent}.

The Casimir effect is usually described
by starting with two closely spaced
parallel metallic plates in a vacuum.
These can be replaced by two closely
spaced concentric spheres to preserve
the spherical symmetry.  According to
Ref. \cite{rG19}, if $a$ is the
magnitude of the separation, then the
pressure $p$ as a function of $a$ is
given by
\begin{equation}\label{E:Casimir1}
   p(a)=-3\frac{\hbar c\pi^2}{720 a^4}
\end{equation}
and the density is
\begin{equation}\label{E:Casimir2}
   \rho_C(a)=-\frac{\hbar c\pi^2}{720a^4}.
\end{equation}
Here $\hbar$ is Planck's constant and $c$
is the speed of light.

At this point we are going to return to
noncommutative geometry by recalling the
form of the energy density in Eq.
(\ref{E:density}) and its interpretation:
the gravitational source causes the mass
$m$ of a particle to be diffused throughout
the region of linear dimension
$\sqrt{\beta}$ due to the uncertainty.
Here we are going to be more concerned
with a smeared spherical surface of which
the throat of a wormhole is our primary
example.  According to Ref. \cite{pK23c},
the energy density $\rho_s$ is given by
\begin{equation}\label{E:rho2}
   \rho_s(r-r_0)=\frac{\mu\sqrt{\beta}}{\pi^2
   [(r-r_0)^2+\beta]^2},
\end{equation}
where $\mu$ now denotes the mass of the
surface.  So the smeared particle is
replaced by a smeared surface.

To connect the Casimir effect to the
noncommutative-geometry background, we
first observe from Eq. (\ref{E:density})
that the energy density $\rho$ as a
function of the separation $a$ is
\begin{equation}
   \rho(a)=\frac{m\sqrt{\beta}}
        {\pi^2(a^2+\beta)^2}.
\end{equation}
According to Eq. (\ref{E:rho2}), in the
vicinity of the throat, i.e., whenever
$r-r_0=a$, we get
\begin{equation}
   \rho_s(a)=\frac{\mu\sqrt{\beta}}
        {\pi^2(a^2+\beta)^2}.
\end{equation}
So it follows from Eq. (\ref{E:Casimir2})
that
\begin{equation}\label{E:Casimir3}
   \frac{\mu\sqrt{\beta}}{\pi^2(a^2+\beta)^2}
   =|\rho_C(a)|=\frac{\hbar c\pi^2}
       {720a^4},
\end{equation}
which is the sought-after connection.  More
precisely, it is argued in Ref. \cite{pK23c}
that the separation $a$, although small, is
still macroscopic. So we can assume that
$\beta=(\sqrt{\beta})^2\ll a^2$.  Moreover,
since $\beta$ is an additive constant, it
becomes negligible in the denominator of
Eq. (\ref{E:Casimir3}); thus
\begin{equation}
   \sqrt{\beta}=\frac{\hbar c\pi^4}
       {720\mu}.
\end{equation}
Since $\hbar=1.0546\times 10^{-34}
\,\text{J}\cdot\, \text{s}$, we obtain
\begin{equation}\label{E:beta}
   \sqrt{\beta}=\frac{4.28\times 10^{-27}}
       {\mu}.
\end{equation}
Given that $\mu$ is the mass of the throat
$r=r_0$, a spherical surface of negligible
thickness, it is hard to quantify, but it
does have a definite value, thereby defining
$\sqrt{\beta}$ in Eq. (\ref{E:beta}).

It is proposed in Ref. \cite{pK23c} that
we could give a direct physical
interpretation to the smearing effect by
letting $\sqrt{\beta}=a$.  Then Eq.
(\ref{E:Casimir3}) yields
\begin{equation}
   \frac{\mu a}{\pi^2(a^2+a^2)^2}=
   \frac{\hbar c\pi^2}{720a^4}
\end{equation}
or
\begin{equation}
   \mu a =\frac{\hbar c\pi^4}{180},
\end{equation}
a fixed quantity.  So there are many
possible choices for $a$ and $\mu$.

Successfully connecting the experimentally
confirmed Casimir effect to noncommutative
geometry has some important consequences.
Here we can follow the arguments proposed
in Ref. \cite{NSS06}, starting with the
assertion that the noncommutative effects
can be implemented in the Einstein field
equations
$G_{\mu\nu}=\frac{8\pi G}{c^4}T_{\mu\nu}$
by modifying only the energy momentum
tensor $T_{\mu\nu}$, while leaving the
Einstein tensor $G_{\mu\nu}$ intact.  The
reason given in Ref. \cite{NSS06} is that
a  metric field is a geometric structure
defined over an underlying manifold whose
strength is measured by its curvature, but
the curvature, in turn,  is nothing more
than the response to the presence of a
mass-energy distribution.
Moreover, the noncommutativity is an
intrinsic property of spacetime, rather
than a superimposed geometric structure.
So it stands to reason that noncommutative
geometry has an effect on the mass-energy
and momentum distributions, which, in
turn, determines the spacetime curvature.
None of this affects  the Einstein tensor.
So the length scales can be macroscopic.
(Recall that we already saw in Sec.
\ref{S:emergent} that the throat radius
 can be macroscopic.)

In summary, by invoking noncommutative
geometry, we have seen that the Casimir
effect, although a small effect, may very
well support a macroscopic wormhole.

\section{Stability}\label{S:stability}
The possible existence of macroscopic
traversable wormholes has naturally led to
numerous studies regarding the stability of
such structures.  We are going to confine
ourselves to Ref. \cite{BHF07} because the
assumption of conformal symmetry in
\cite{BHF07} can be combined with the
noncommutative-geometry background to
produce a complete wormhole solution.  It
is assumed in Ref. \cite{BHF07} that our static
and spherically symmetric spacetime admits
a one-parameter group of conformal motions.
This assumption is equivalent to the
existence of conformal Killing vectors such
that
\begin{equation}
   \mathcal{L_{\xi}}g_{\mu\nu}=g_{\eta\nu}\,\xi^{\eta}
   _{\phantom{A};\mu}+g_{\mu\eta}\,\xi^{\eta}_{\phantom{A};
   \nu}=\psi(r)\,g_{\mu\nu},
\end{equation}
where the left-hand side is the Lie derivative
of the metric tensor and $\psi(r)$ is the
conformal factor \cite{BHF07}.  According to
the usual terminology, $\xi$ generates the
conformal symmetry and the metric tensor
$g_{\mu\nu}$ is said to be conformally mapped
into itself along $\xi$.  This type of
symmetry has been used extensively in classical
general relativity.

Before returning to the stability question,
we need to recall the usual strategy in the
theoretical construction of a Morris-Thorne
wormhole: retain complete control over the
geometry by specifying the redshift and shape
functions and then manufacture or search
the Universe for materials or fields that
produce the required stress-energy tensor.
Ref. \cite{pK16} addresses this problem
in a direct manner: the
noncommutative-geometry background
produces the shape function and the
conformal symmetry yields the redshift
function.  Adding the assumption of the
conservation of mass-energy then yields
the stress-energy tensor.  The result
is a complete wormhole solution determined
from the given conditions.  Finally, it
is shown that the wormhole is stable to
linearized radial perturbations.

\section{Other low energy-density wormholes}
   \label{S:other}
Returning to Eq. (\ref{E:null2}), we have
seen that the local violation of the NEC
can be viewed as a fundamental property in a
noncommutative-geometry setting from which
emerges the violation on a macroscopic
scale.  This observation is consistent with
Eq. (\ref{E:Ein1}), restated here for
convenience:
 \begin{equation}\label{E:restate}
    \frac{b'(r)}{8\pi r^2}=\rho(r);
 \end{equation}
the left-hand side is the $G_{tt}$
component of the Einstein tensor, which,
as we saw in Section \ref{S:Casimir}, is
unaffected in a noncommutative-geometry
setting.  So we are justified in using
$\rho(r)$ from Eq. (\ref{E:density})
in Eq. (\ref{E:restate}).  It is
interesting to note that the small value
of $\rho(r)$ typically leads to $b'(r)<1$.
So the flare-out condition is met
automatically.

The real question now becomes, what if
$\rho(r)$ has a small value but is
otherwise arbitrary?  While we still have
$b'(r)<1$, without the special features
from the noncommutative-geometry
background, we cannot simply and
uncritically draw the same conclusions.
To see why, consider an extreme example,
the zero-density case $\rho\equiv 0$,
treated in Visser's book \cite{mV95}.
We get a valid wormhole solution only
if we go back to requiring the usual
exotic matter.  Unfortunately,
similar comments can be made about
various dark-matter models all of
which have a very low energy density.
Consider, for example, the
Navarro-Frenk-White model in Ref.
\cite{RKRI},
\begin{equation}\label{E:NFW}
   \rho(r)=\frac{\rho_s}{\frac{r}{r_s}
   {\left(1+\frac{r}{r_s}\right)^2}},
\end{equation}
where $r_s$ is the characteristic scale
radius and $\rho_s$ the corresponding
density.  Since $\rho(r)$ in Eq.
(\ref{E:NFW}) is very small, we would
normally satisfy the flare-out condition
$b'(r)<1$ for all $r$, thereby yielding
a wormhole solution.  However, without
the noncommutative-geometry background
and its many special properties, we
can no longer assume the validity of
Eq. (\ref{E:restate}) for an arbitrary
$\rho(r)$, unless, of course, we return
to the exotic-matter requirement once
again.  While this outcome does not
invalidate the solutions, it does call
into question their relevance: if exotic
matter is needed anyway, then what is the
role of dark matter, if any?  In other
words, if exotic matter cannot be
eliminated, then dark matter alone could
not support traversable wormholes.  The
same comments apply equally well to
dark-energy models that do not cross the
phantom divide.  To clarify this point,
we need to recall that for phantom dark
energy, the (isotropic) equation of state
is $p=\omega\rho$, $\omega<-1$, which
implies that $\rho +p=\rho +\omega\rho=
\rho(1+\omega)<0$.  Since the NEC has been
violated, phantom dark energy could in
principle support traversable wormholes
\cite{sS05}.  Such wormholes could only
exist on very large scales, however, as
we already noted in Sec. \ref{S:tension}.

\section{A solution uncovered via
   gravitational lensing}\label{S:lensing}

We know from the previous section that
neither dark matter alone nor dark energy
alone can support a Morris-Thorne wormhole:
the former requires the existence of exotic
matter and the latter the equation of state
$p=\omega\rho$, $\omega <-1$.  Another
possibility is a noncommutative-geometry
background, as we saw in Sec.
\ref{S:emergent}.  This section considers
yet another approach, discussing the effect
of gravitational lensing.  While primarily
a tool for detecting wormholes, it has its
own physical requirements, as described in
Ref. \cite{pK14}.  To facilitate the
discussion, the line element is written
in the following more convenient form:
\begin{equation}
ds^{2}=-A(x)\,dt^{2}+B(x)\,dx^2
+C(x)(d\theta^{2}+\text{sin}^{2}\theta\,
d\phi^{2}),
\end{equation}
where $x$ is the radial distance defined
in terms of the Schwarzschild radius
$x=r/2M$.  Then
\begin{equation}
   x_0=\frac{r_0}{2M}
\end{equation}
denotes the closest approach of the
light ray.  As noted in Ref. \cite{pK14},
the deflection angle $\alpha(r_0)$ is
given by
\begin{equation}
   \alpha(r_0)=I(x_0)+a,
\end{equation}
where $a$ is a constant that depends on
the size of the wormhole.  Next,
\begin{equation}
   I(x_0)=2\int^{\infty}_{x_0}\frac
   {\sqrt{B(x)}\,dx}{\sqrt{C(x)}
   \sqrt{\frac{C(x)A(x_0)}{C(x_0)A(x)}}-1}
   =\int^a_{x_0}Q(x)\,dx.
\end{equation}
Here $Q(x)$ depends on the parameters
in the Navarro-Frenk-White model, Eq.
(\ref{E:NFW}).  (See Ref. \cite{pK14}
for details.)

So while we are still dealing with dark
matter, the wormhole solution requires
several other conditions besides the
simple existence of dark matter as
previously claimed.  In particular,
the deflection angle depends on both
the redshift and shape functions.

\section{Conclusion}

Given that wormholes are just as good a
prediction of Einstein's theory as black
holes, we can assume that Morris-Thorne
wormholes, as proposed in Ref. \cite{MT88},
are theoretically possible, but subject to
severe restrictions from quantum field
theory.  The purpose of this paper is to
show that these restrictions can be met
without a modification of Einstein's
theory.

Adhering to the widely held view that the
need for exotic matter renders any wormhole
solution unphysical, we follow Ref.
\cite{pK23a} which proposes the following
static and spherically symmetric line
element to describe a  wormhole spacetime:
\begin{equation*}
ds^{2}=-e^{2\Phi(r)}dt^{2}+\frac{dr^2}{1-\frac{b(r)}{r}}
+r^{2}(d\theta^{2}+\text{sin}^{2}\theta\,
d\phi^{2})+e^{2\mu(r,l)}dl^2,
\end{equation*}
where $l$ is the extra coordinate.  The extra
dimension can be either static or time
dependent.  For this approach to work, we do
need to make some additional assumptions.  For
the static case, the redshift function is
positive and decreasing for all $r$.
(Asymptotic flatness is part of the structure
of a Morris-Thorne wormhole.)  Furthermore,
Inequalities (\ref{E:condition1}) and
(\ref{E:condition3}) have to be met, while
the condition $b'(r_0)>\frac{1}{3}$ ensures
that the NEC is satisfied at the throat for
all null vectors.  So the throat of the
wormhole can be lined with ordinary matter,
while the unavoidable violation of the NEC
can be attributed to the higher spatial
dimension.  Finally, the extra dimension
can be small or even curled up.

For the time-dependent case, we obtain a
similar conclusion using the slightly more
general condition
$\partial\mu(r_0,l,t)/\partial r\le -2/r_0$,
provided that $a''(t)<0$.

The next part of this paper invokes a
noncommutative-geometry background, thereby
assuming that point-particles are replaced
by smeared objects, as detailed in Sec.
\ref{S:emergent}.  This assumption is
consistent with the Heisenberg uncertainty
principle and therefore independent of
Einstein's theory.  The particle-like
gravitational source is given by
\begin{equation}
   \rho(r)=\frac{m\sqrt{\beta}}{\pi^2
   (r^2+\beta)^2}.
\end{equation}
So the gravitational source causes the
mass $m$ of the particle to be diffused
throughout the region of linear dimension
$\sqrt{\beta}$ due to the uncertainty.

The shape function $B$ meets all the usual
requirements; in particular,
\begin{equation}
   B\left(\frac{r_0}{\sqrt{\beta}}\right)=
   \frac{r_0}{\sqrt{\beta}}.
 \end{equation}
 The throat radius $r_0/\sqrt{\beta}$ is
 therefore macroscopic.

It is argued in Sec. \ref{S:emergent}
that the noncommutative-geometry background
is a \emph{fundamental property} and the
outcome, a macroscopic wormhole, is an
\emph{emergent phenomenon}.  The result
is an \emph{effective model} that does
not depend on the short-distance effect
that is characteristic of noncommutative
geometry, thereby avoiding a modification
of Einstein's theory.  It is interesting
to note that the compactified extra spatial
dimension in Sec. \ref{S:extra} can also
be viewed as a fundamental property,
again making the macroscopic wormhole
an emergent phenomenon.

It was pointed out in Sec. \ref{S:tension}
that for a wormhole with a throat radius
of 3 km, the radial tension has the same
magnitude as the pressure at the center
of a massive neutron star, suggesting that
a Morris-Thorne wormhole should be viewed
as a compact stellar object.  This case
is taken up in Sec. \ref{S:neutron} by
first noting that quark matter is believed
to exist near the center of neutron stars,
thereby calling for a combined model
consisting of quark matter and ordinary
matter.  For this type of wormhole, the
violation of the null energy condition
can be attributed to the extreme
conditions at the center of the neutron star.

In Sec. \ref{S:finetune} we saw return to
exotic matter, motivated in part by the
fact that small amounts of exotic matter
can be made in the laboratory, as exemplified
by the Casimir effect.  To get a valid
wormhole solution, the wormhole has to
satisfy the Ford-Roman Inequality
(\ref{E:FR}) or the extended version,
Inequality (\ref{E:pK}).  The conclusion
is that one must strike a balance between
the thickness $[r_0,r_1]$ of the exotic
region and the degree of fine-tuning
required to achieve this reduction.  The
degree of fine-tuning is a generic feature
of a Morris-Thorne wormhole.  Sec.
\ref{S:Casimir} then connects the
aforementioned Casimir effect with
noncommutative geometry, suggesting that
the former may be able to support a
macroscopic wormhole in spite of being a
small effect.

Finally, it is shown in Sec. \ref{S:stability}
that given a noncommutative-geometry background,
a Morris-Thorne wormhole in a static and
spherically symmetric spacetime admitting
conformal motion is stable to linearized
radial perturbations.  Furthermore, the
redshift and shape functions are completely
determined from the given conditions.

In Sec. \ref{S:other} we return to Eq.
(\ref{E:restate}) to observe that $b'(r)=
8\pi r^2\rho(r)<1$ whenever $\rho(r)$ is
extremely small, which is actually true in
a dark-matter or dark-energy setting, as
well as for the zero-density case $\rho(r)
\equiv 0$.  So the NEC is automatically
violated.  The same is true for $\rho(r)$
in Eq. (\ref{E:density}), the
noncommutative-geometry case.  The
difference is that the use of $\rho(r)$
in Eq. (\ref{E:density}) can be justified
by appealing to the special properties of
the noncommutative-geometry background,
thereby producing a valid wormhole solution.
Since these key properties are not possessed
by any of the other cases, we conclude that
neither dark matter nor dark energy can
support a Morris-Thorne wormhole, as long
as the latter does not cross the phantom
divide.  Other possible exceptions are
noted in Ref. \cite{pK14}.

Sec. \ref{S:lensing} discusses the detection
of wormholes by means of gravitational
lensing.  An application of the method calls
for additional physical requirements beyond
the simple existence of dark matter, confirming
the earlier assertion that dark matter alone
cannot support traversable wormholes.
\\
\\
\noindent
CONFLICTS OF INTEREST
\\
\noindent
The author declares that there are no
conflicts of interest regarding the
publication of this paper.

\end{document}